\documentclass[a4paper]{jpconf}
\usepackage{graphicx}
\usepackage{tikz}
\usepackage{amsmath}

\begin{document}
	\title{Turbulent activity in the near-wall region of adverse pressure gradient turbulent boundary layers}
	
	\author{Taygun R Gungor$^{1}$, Yvan Maciel$^1$, Ayse G Gungor$^2$}
	
	\address{$^1$ Department of Mechanical Engineering, Universit\'e Laval, Quebec City, QC, G1V 0A6 Canada \\ $^2$ Faculty of Aeronautics and Astronautics, Istanbul Technical University, 34469 Maslak, Istanbul, Turkey }
	
	\ead{taygun-recep.gungor.1@ulaval.ca}

	\newcommand{\mi}{\boldsymbol{-} \mathrel{\mkern -14mu} \boldsymbol{-}}

	\begin{abstract}
		
		Two direct numerical simulation (DNS) databases are investigated to understand the effect of the outer-layer turbulence on the inner layer's structures and energy transfer mechanisms. The first DNS database is the non-equilibrium adverse-pressure-gradient (APG) turbulence boundary layer (TBL) of Gungor et al. \cite{gungor2022energy}. Its Reynolds number and the inner-layer pressure gradient parameter reach above 8000 and 10, respectively. The shape factor spans between 1.4 and 3.3, which indicates the flow has various velocity defect situations. The second database is the same flow as the first one but the outer layer turbulence is artificially eliminated in this flow. Turbulence is removed above 0.15 local boundary layer thickness. For the analysis, we chose four streamwise positions with small, moderate, large, and very-large velocity defect. We compare the wall-normal distribution of Reynolds stresses, two-point correlations and spectral distributions of energy, production and pressure strain. The results show that the inner layer turbulence can sustain itself when the outer-layer turbulence does not exist regardless of the velocity defect or the pressure gradient. The two-point correlations of both cases show that outer large-scale structures affect the inner layer structures significantly. The streamwise extent of the correlation contours scales with pressure-viscous units. This shows the importance of the pressure gradient's effect on the inner-layer structures. The spectral distributions demonstrate that the energy transfer mechanisms are probably the same in the inner layer regardless of the velocity defect, which suggests the near-wall cycle may exist even in very-large defect APG TBLs where the mean shear in the inner layer is considerably lower than small-defect APG TBLs.

	\end{abstract}

	\section{Introduction}
	
	When a turbulent boundary layer (TBL) is exposed to an adverse pressure gradient (APG), it decelerates. A sustained and/or strong deceleration leads the mean velocity profile and consequently the mean shear profile to change dramatically. The mean shear increases significantly in the outer layer. As a result of this, the outer layer turbulence becomes dominant and an outer maximum for the streamwise Reynolds normal stress ($\langle u^2\rangle$) emerges when the defect is large enough \cite{skaare1994turbulent, gungor2016scaling}. In contrast, the mean shear is attenuated in the inner layer and therefore the inner-layer turbulent activity becomes weaker and loses its importance \cite{skaare1994turbulent, elsberry2000experimental}. An important outcome is that the well-known inner-peak of the streamwise Reynolds stresses vanishes in large defect APG TBLs \cite{maciel2006study,gungor2016scaling}.

	In the inner layer, the turbulent structures are not affected heavily by the effect of APG when the defect is small \cite{lee2017large,harun2013pressure}. The spectral distributions of energy are very similar in canonical flows and small-defect APG TBLs \cite{kitsios2017direct,gungor2022energy}. Contrary to this, the spatial features and spectral characteristics of structures become different in large-defect APG TBLs. The near-wall streaks are irregular and less streaky and furthermore, the sweeps/ejections are weaker in large-defect APG TBLs \cite{skote2002direct, maciel2017structural,maciel2017coherent}. Moreover, the inner peak in the $\langle u^2\rangle$-spectra does not exist \cite{lee2017large,gungor2022energy}. The near-wall cycle may therefore be modified or hidden among the footprints of the large-scale outer-layer structures that are energetic in large-defect TBLs.
	
	Indeed, inner-layer structures in large-defect APG TBLs are affected heavily by the outer-layer energetic structures. The one-dimensional energy spectra show that most of the energy of the $u$-structures in the inner layer resides at the same wavelengths as outer-layer large-scale structures \cite{kitsios2017direct,gungor2022energy}. Gungor et al. \cite{gungor2022energy} demonstrated that even the two-dimensional energy spectra do not show any significant energy at wavelengths that are associated with the inner layer streaks (streamwise wavelength $\lambda_x^+\approx 1000$ and spanwise wavelength $\lambda_z^+\approx 120$, where superscript $^+$ refers to friction-viscous units). The most energetic structures in the inner-layer have very wide wavelengths ($\lambda_z^+\approx 500$), which is very similar to outer-layer energetic structures. These findings suggest that the inner layer is dominated by imprints of large-scale outer-layer structures.

	Besides the structures' characteristics, the existence of the near-wall cycle in large-defect APG TBLs is another crucial topic to address. To this date, however, there has been no conclusive evidence of whether the near-wall cycle in large-defect APG TBLs exists or not. Unfortunately, only a few studies focused on this issue. Jiao et al. \cite{jiao2022driving} suggested that the classical near-wall cycle is weakened heavily or dissolved completely under the effect of APG. They found a new mechanism of near-wall turbulence production initiated by the wall-normal nonlinear transport of the outer wall-normal velocity fluctuations to the near-wall region. It is important to highlight that their study is based on a Couette flow and not a TBL. The inner-layer pressure-gradient parameter ($\beta_i=(\nu/(\rho u_	\tau^3))(d P_w/dx)$, where $P_w$ is the pressure at the wall, $\nu$ is viscosity and $\rho$ is density) reaches up to $0.137$ in their case but the Reynolds number is $Re_\tau\approx 57$ when $\beta_i=0.137$. In contrast to this, Gungor et al. \cite{gungor2022energy} suggested that the near-wall cycle still exists in large-defect APG TBLs even though the spectral signature of the streaks disappears. They hypothesized that the near-wall cycle or another self-sustaining process with similar features may exist without the streaks. Even in canonical flows, Jim\'enez \cite{jimenez2022streaks} showed later that the near-wall cycle was possible without streaks and that the latter might be byproducts of the bursting process. Moreover, Gungor et al. \cite{gungor2022energy} claimed that the inner-layer streaks may be still there but we simply cannot see or detect them because the large-scale outer-layer structures that dominate the flow in the inner layer hide the streaks and streaks' signature.

The present study aims to investigate the effect of the outer-layer turbulence on the inner-layer turbulent activity in APG TBLs. For this, we employ two direct numerical simulation (DNS) databases. The first one is a non-equilibrium APG TBL taken from the literature \cite{gungor2022energy}. Its Reynolds number based on momentum thickness ($Re_\theta$) and $\beta_i$ reaches above 8000 and 10, respectively. The shape factor spans between 1.4 and $3.3$. The second one is a novel non-equilibrium APG TBL designed based on the first one. This database's most important feature is that the outer-layer turbulence is artificially removed to investigate the inner layer without the effect of the outer layer. We compared the wall-normal distribution of Reynolds stresses, the two-point correlations and the spectral distributions of energy, production, and pressure-strain of these two cases.

	\section{Databases}

	\begin{figure}

\centering
		\includegraphics[scale=0.45]{ 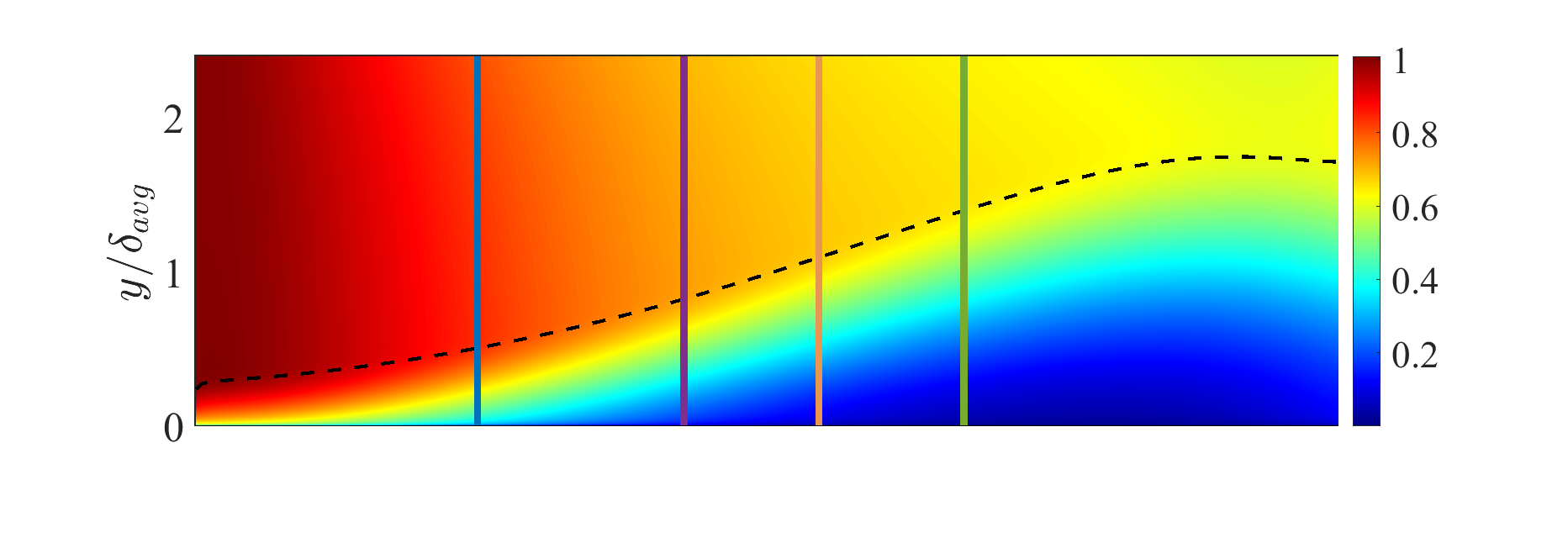} \vspace{-0.8cm}

		\includegraphics[scale=0.45]{  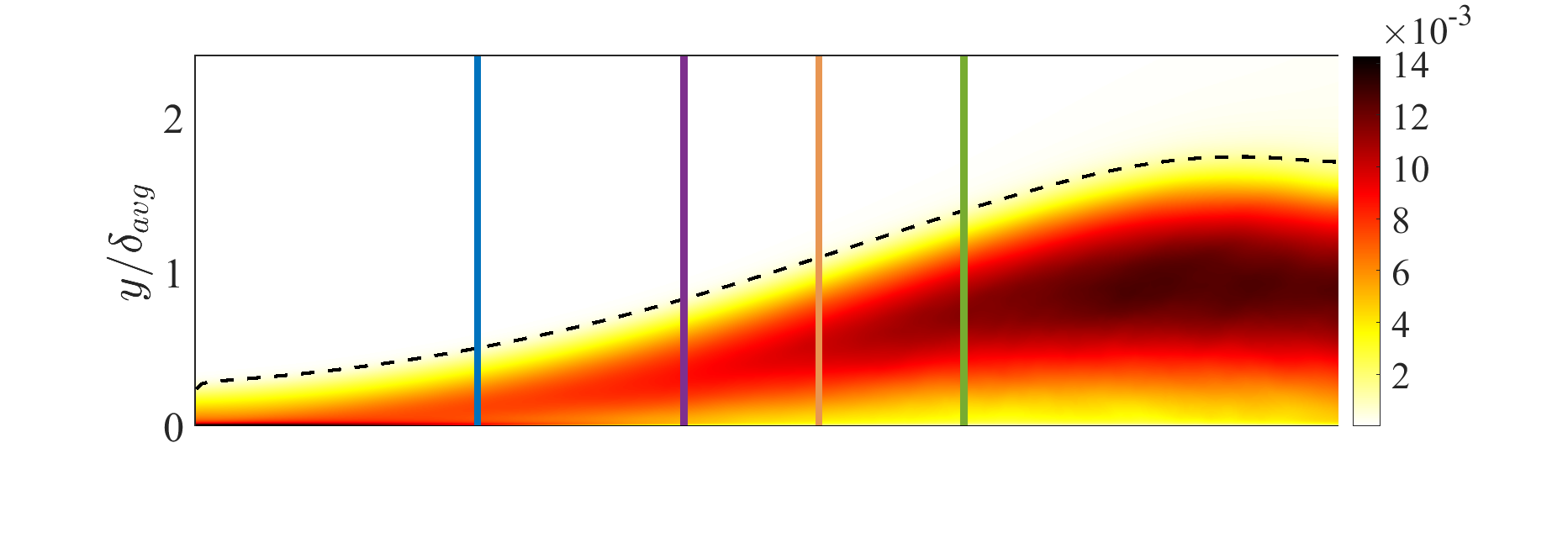} \vspace{-0.8cm} 
		
		\includegraphics[scale=0.45]{ 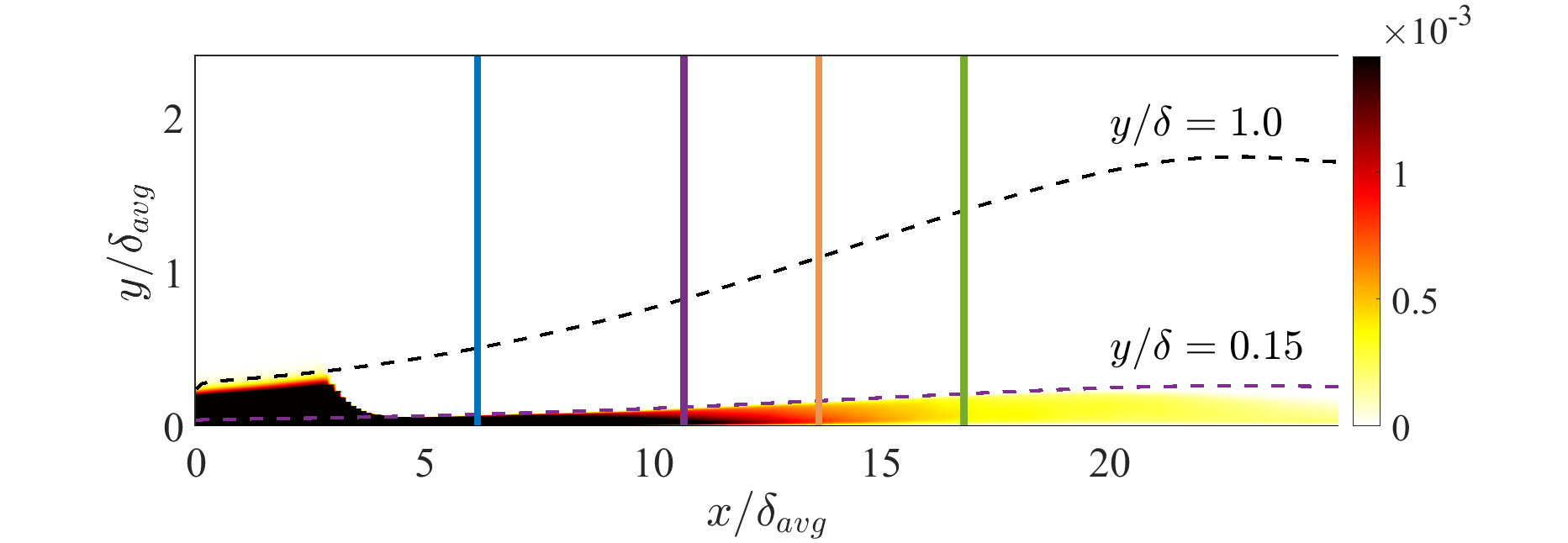}
		\caption{The streamwise distribution of the mean flow (top) and $\langle u^2\rangle$ of the original case (middle) and the manipulated case (bottom) as a function of $x$ and $y$. The axes are normalized with $\delta_{avg}$ and the levels are normalized with the edge velocity at the inlet. The straight lines represent the four streamwise positions (SP1 to SP4). The dashed lines indicate $\delta$ and $0.15\delta$. Only a part of the domain in the wall-normal direction is shown.} 
		\label{mean_flow_and_u2}
	\end{figure}

	The first database, generated using DNS, is described in detail in Ref.\cite{gungor2022energy}. As mentioned above, it is a non-equilibrium APG TBL that is exposed to a strong APG. At the beginning of the domain, the flow is a ZPG TBL but the effect of the pressure gradient leads to an increasing defect in the mean velocity profile. Towards the end of the domain, the flow evolves into a TBL at the verge of separation due to the APG. The Reynolds number based on momentum thickness $(Re_\theta)$ and the shape factor ($H$) span between 1400-8200 and 1.4-3.3, respectively and $\beta_i$ reaches above 10. Figure \ref{mean_flow_and_u2}($a,b$)  shows the mean velocity and $\langle u^2\rangle$ as a function of $x$ and $y$, where the axes are normalized with the boundary layer thickness averaged in the region of interest ($\delta_{avg}$). The mean velocity distribution shows that the boundary layer decelerates and the boundary layer thickness ($\delta$) grows in height as the flow develops. As for the turbulence, $\langle u^2\rangle$ is high in the inner layer at the beginning of the domain. But this changes and the maximum turbulence is found in the outer layer as the defect increases.

	The second APG TBL database is a manipulated APG TBL generated using the same DNS case mentioned above. The flow is basically the same as the previous one with a major exception. The outer layer turbulence is artificially removed above the inner layer in the zone that corresponds to the wake layer in canonical wall flows, that is above the overlap layer. There is no commonly accepted layer definition for APG TBLs \cite{maciel2018outer}. Therefore, we have assumed the inner layer, the zone where viscous effects are important, to be between the wall and $0.15\delta$, where $\delta$ is the local boundary thickness. The fluctuations and therefore the Reynolds stresses are zero in the region above $0.15\delta$ as will be discussed later in the paper. 
	
	 The DNS code utilizes spectral discretization in the homogeneous spanwise direction (More information on the code and or the DNS setup can be found in references \cite{simens2009high, borrell2013code,gungor2016scaling}). Therefore, the instantaneous flow field is stored as a function of $x$, $y$ and $k_z$ where $k_z$ is the wavenumber in the spanwise direction. To eliminate turbulence in the outer layer, all spanwise modes with the exception of the zeroth mode, which is the mean flow, are set to zero at each time step as follows

	\begin{equation}
		\widehat{U}_i (x,y,k_z ; y>0.15\delta, k_z=1:n_z) = 0,
		\label{removefluc}
	\end{equation}

	\noindent where $n_z$ is the number of spanwise modes and $\hat{U}_i$ is the Fourier transform of the velocity signal. The issue with eliminating turbulence is that once the turbulence is removed, the mean velocity changes significantly. However, we want to prevent this from happening because we want to keep both flows identical to isolate the effect of outer-layer turbulence as much as possible. To keep both flows identical, the mean velocity (averaged in time and spanwise direction) from the original case is imposed to the zeroth mode of the velocity of the manipulated case at each time step as

	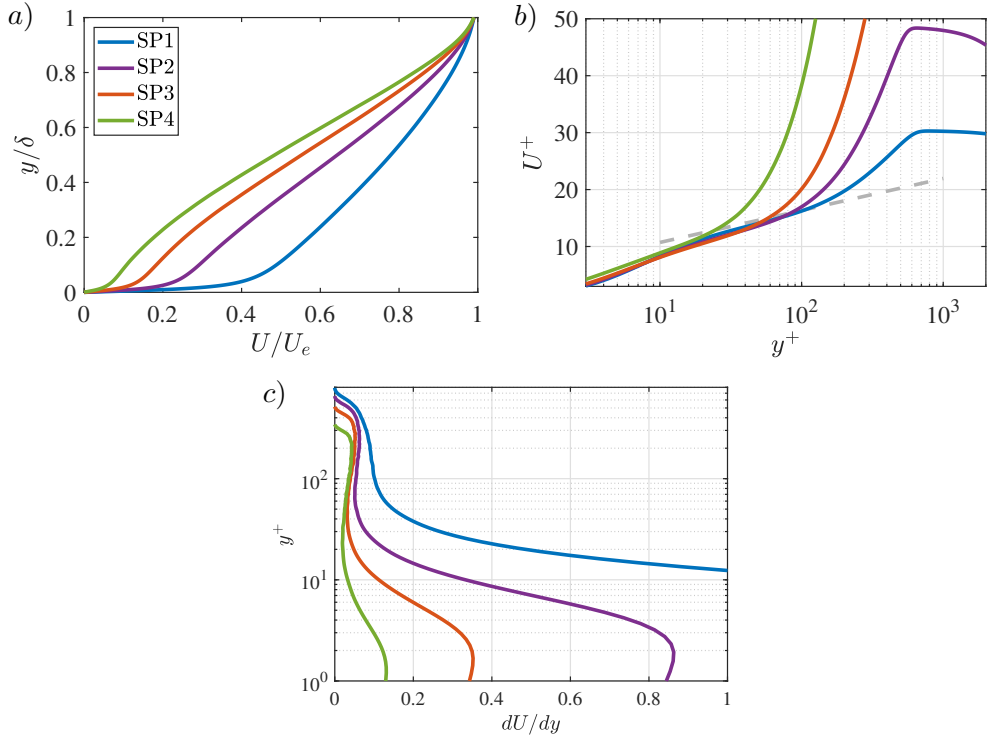
\begin{figure}
		\centering

\begin{tikzpicture}
\node(a){ \includegraphics[scale=0.45]{ 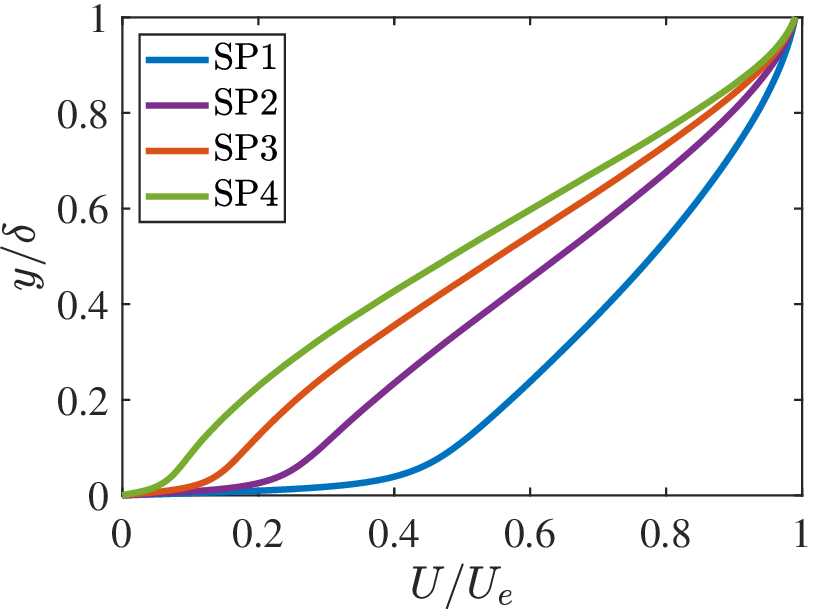}};
	{\node at (-3,2.2) {  $a)$};}
\end{tikzpicture} \begin{tikzpicture}
\node(a){ \includegraphics[scale=0.45]{  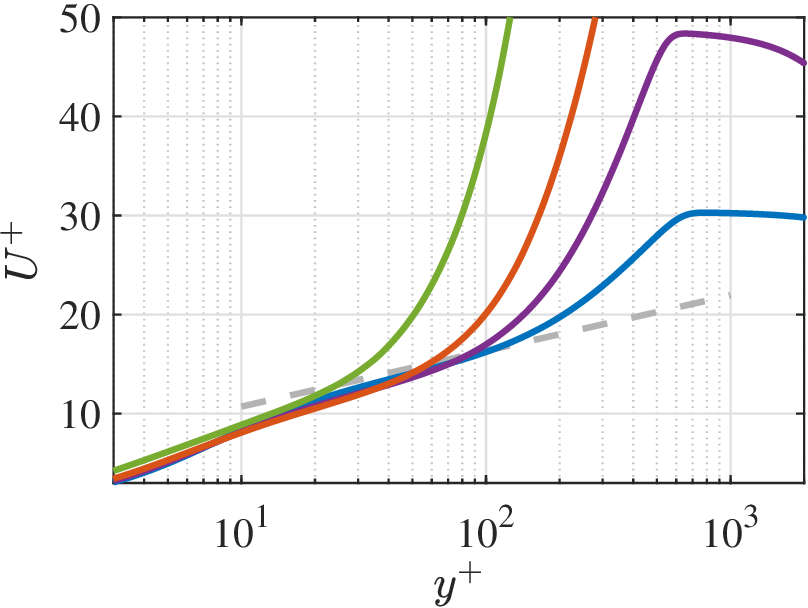}};
	{\node at (-3,2.2) {  $b)$};}
\end{tikzpicture}

\begin{tikzpicture}
\node(a){ \includegraphics[scale=0.45]{  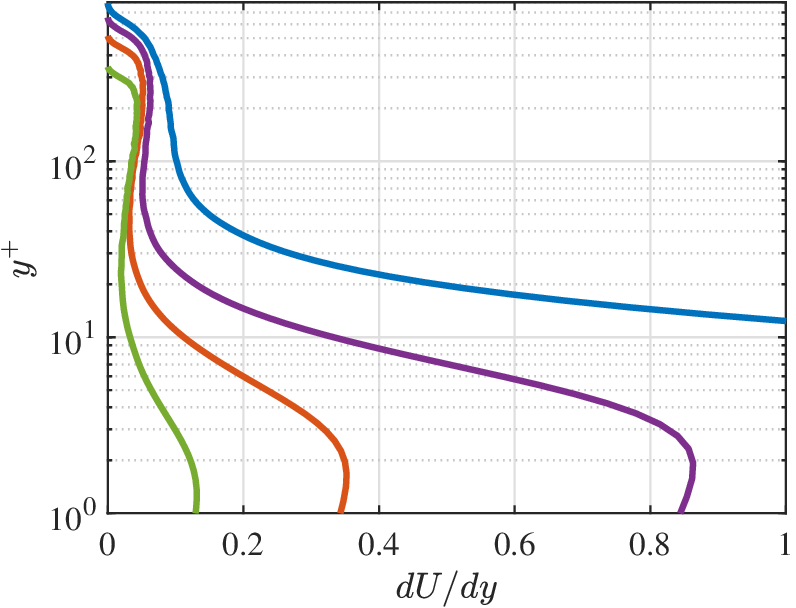}};
	{\node at (-3,2.2) {  $c)$};}
\end{tikzpicture}		
					
		\caption{The mean velocity profiles as a function of $y$. The levels and axes are normalized with outer scales ($a$) and friction-viscous scales ($b$). The gray line is the logarithmic law with $\kappa=0.41$ and $B=5.1$. The mean shear profile as a function of $y^+$ ($c$). The mean shear is not normalized.}
		\label{mean_vel}
	\end{figure} 
	
	\begin{equation}
		\widehat{U}_i (x,y,k_z ; k_z=0) = U_{i,original}(x,y).
		\label{zerothmode}
	\end{equation}
	
	\noindent Figure \ref{mean_flow_and_u2}(c) shows the spatial evolution of $\langle u^2\rangle$ in the manipulated case as a function of $x/\delta_{avg}$ and $y/\delta_{avg}$. The distribution shows the region where the outer layer turbulence activity is removed. The elimination of turbulence starts at approximately $3\delta_{avg}$ at the inlet to avoid some numerical issues. 
	
	We choose four streamwise positions for comparison of these two cases, as shown with straight lines in figure \ref{mean_flow_and_u2}. The main parameters of these four stations are given in Table 1. These positions correspond to small, moderate, large and very-large velocity defect situations. Figure \ref{mean_vel}($a,b$) presents the outer-scaled and inner-scaled mean velocity profiles as a function of wall-normal distance ($y$). The outer-scaled mean profiles demonstrate the dramatic changes in the mean velocity. Whereas the profile at SP1 resembles canonical wall flows, the profiles progressively change with increasing velocity defect. The inner-scaled profiles show that none of the flows follow the traditional logarithmic law. Despite this, there is a clear trend that the deviation from the logarithmic law increases with increasing velocity defect \cite{gungor2022energy}. Figure \ref{mean_vel}($c$) presents the mean shear profiles without any normalization because there are not appropriate scales for mean shear levels that we can use for both small- and large-defect APG TBLs. As the profiles indicate, the mean shear in the inner layer is inversely proportional to the velocity defect.

	\begin{table}
		\centering
		\label{Table1}
		\caption{The main properties of the streamwise positions}		\begin{tabular}{c  c  c   c  c  c }
			& & \\ 
			\hline
			\hline
			
			Name	& $Re_\theta$	& $Re_\tau$	&$H$ & $C_f$  & $\beta_i$  \\
			\hline
			SP1o/SP1m	& 3004	& 645 &  1.65& 0.00217&  0.0278  \\ 
			SP2o/SP2m	& 4740	& 574 &  2.14& 0.00087&  0.0831  \\
			SP3o/SP3m	& 5796	& 458 &  2.63& 0.00035&  0.1928  \\ 
			SP4o/SP4m	& 6734	& 340 &  3.11& 0.00013&  0.6184  \\  
			SPH293  	& 6342  & 383 &  2.91& 0.00019&  0.3832  \\  
			SPH300 		& 6466  & 356 &  2.97& 0.00016&  0.5049  \\  
			\hline
			\hline
		\end{tabular}

	\end{table}

	\begin{figure}[h!]
		\centering
		
\begin{tikzpicture}
\hspace*{-0.85cm}	\node(a){ \includegraphics[scale=0.5]{  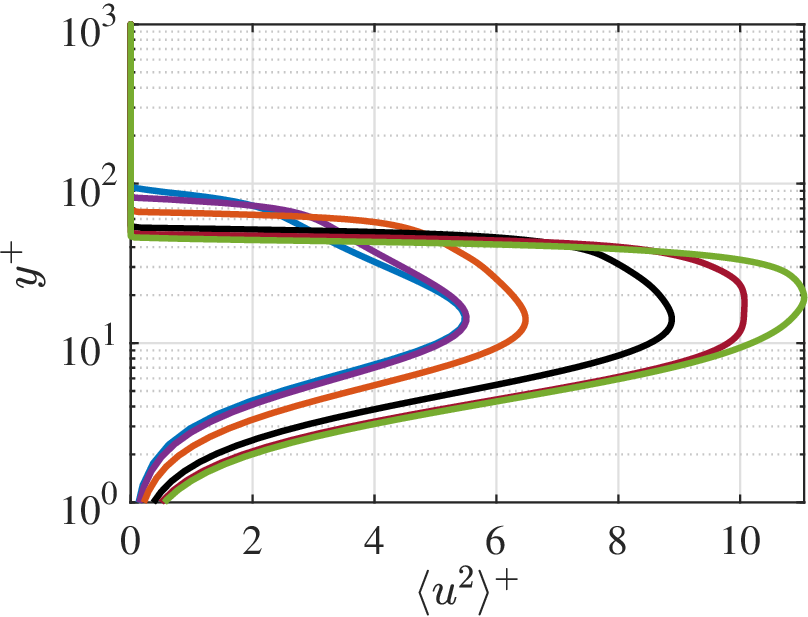}};
	{\node at (-3.3,2.3) {  $a)$};}
\end{tikzpicture}
\begin{tikzpicture}
\hspace*{-1.1cm}	\node(a){ \includegraphics[scale=0.5]{  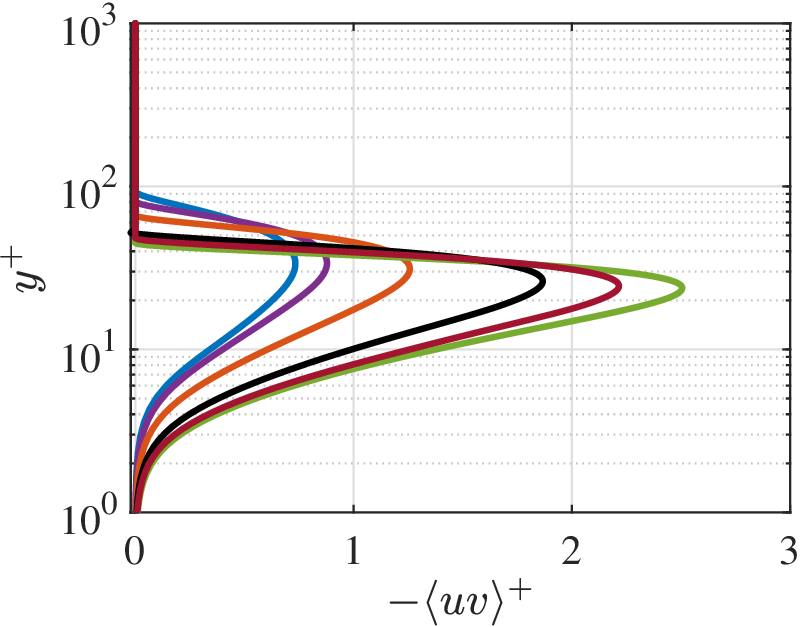}};
	{\node at (-3.3,2.3) {  $b)$};}
\end{tikzpicture} 

\begin{tikzpicture}
\hspace*{-0.55cm} \node(a){ \includegraphics[scale=0.5]{ 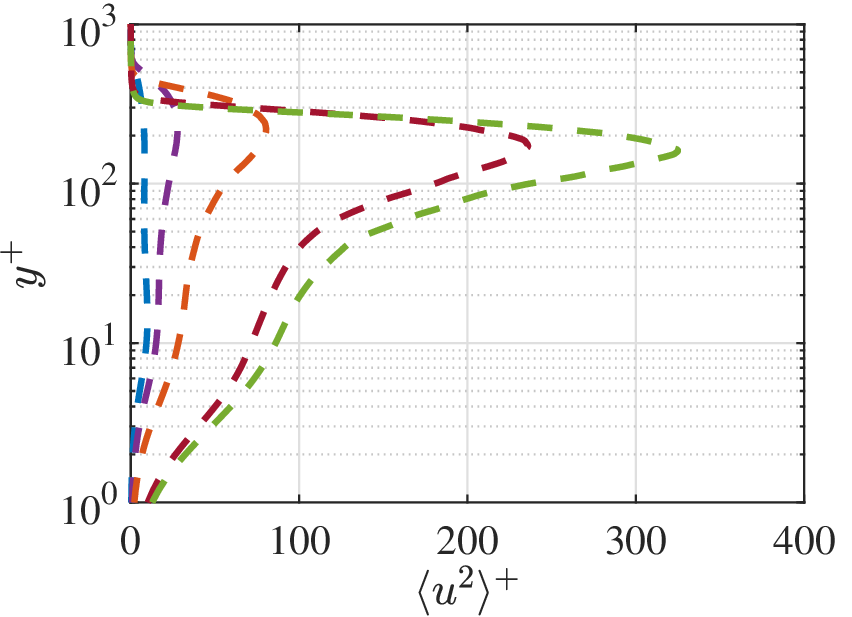}};
	{\node at (-3.3,2.3) {  $c)$};}
\end{tikzpicture}
\begin{tikzpicture}
\hspace*{-1cm} \node(a){ \includegraphics[scale=0.5]{  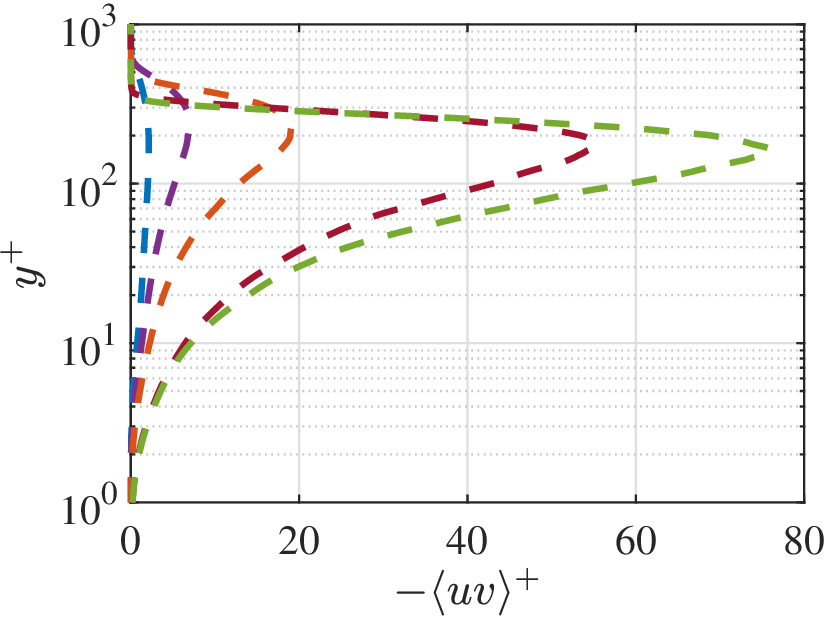}};
	{\node at (-3.3,2.3) {  $d)$};}
\end{tikzpicture}

\begin{tikzpicture}
	\node(a){ \includegraphics[scale=0.5]{  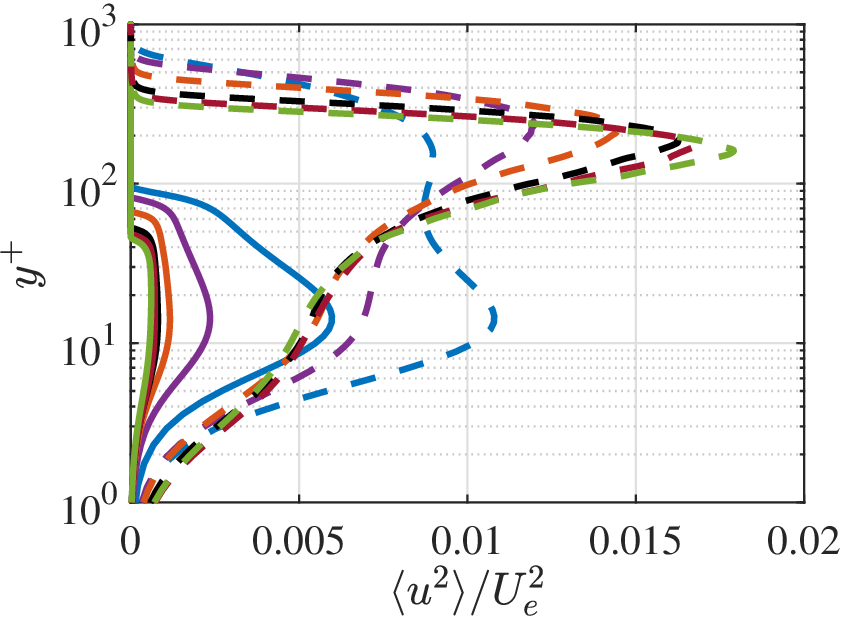}};
	{\node at (-3.3,2.3) {  $e)$};}
\end{tikzpicture}\begin{tikzpicture}
\hspace*{-.25cm}	\node(a){ \includegraphics[scale=0.5]{  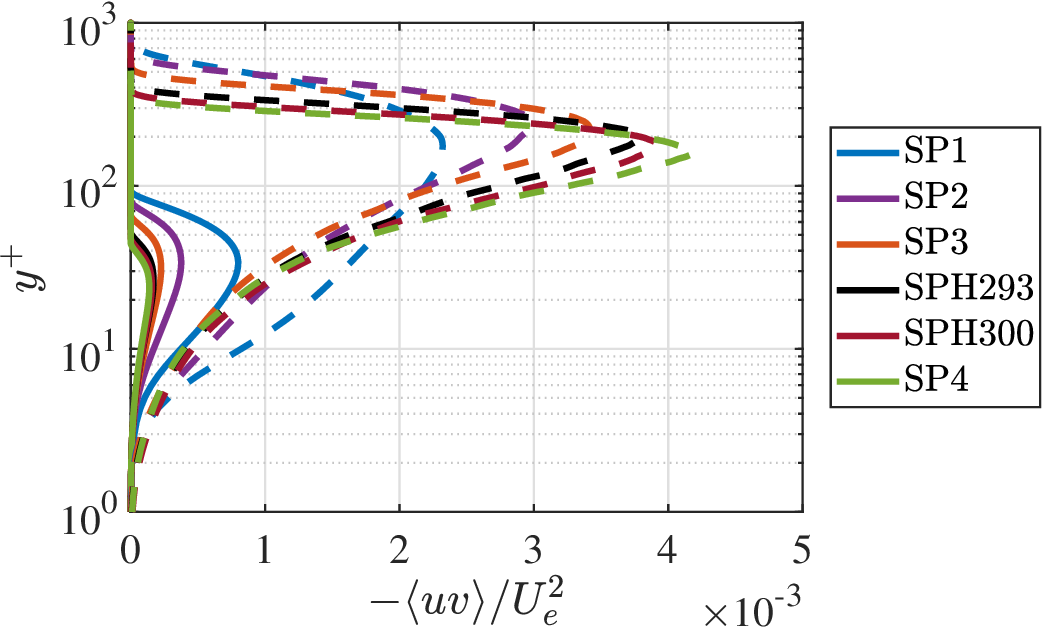}};
	{\node at (-4.2,2.3) {  $f)$};}
\end{tikzpicture}

\caption{The $\langle u^2\rangle^+$ and $\langle uv\rangle ^+$ profiles ($a,b,c,d$) and the $\langle u^2\rangle/U_e^2$ and $\langle uv\rangle/U_e^2$ profiles ($e,f$) as a function of $y^+$ for the four streamwise positions. The dashed and straight lines are for the original and manipulated cases, respectively. } 
		\label{rs_profiles}
	\end{figure}
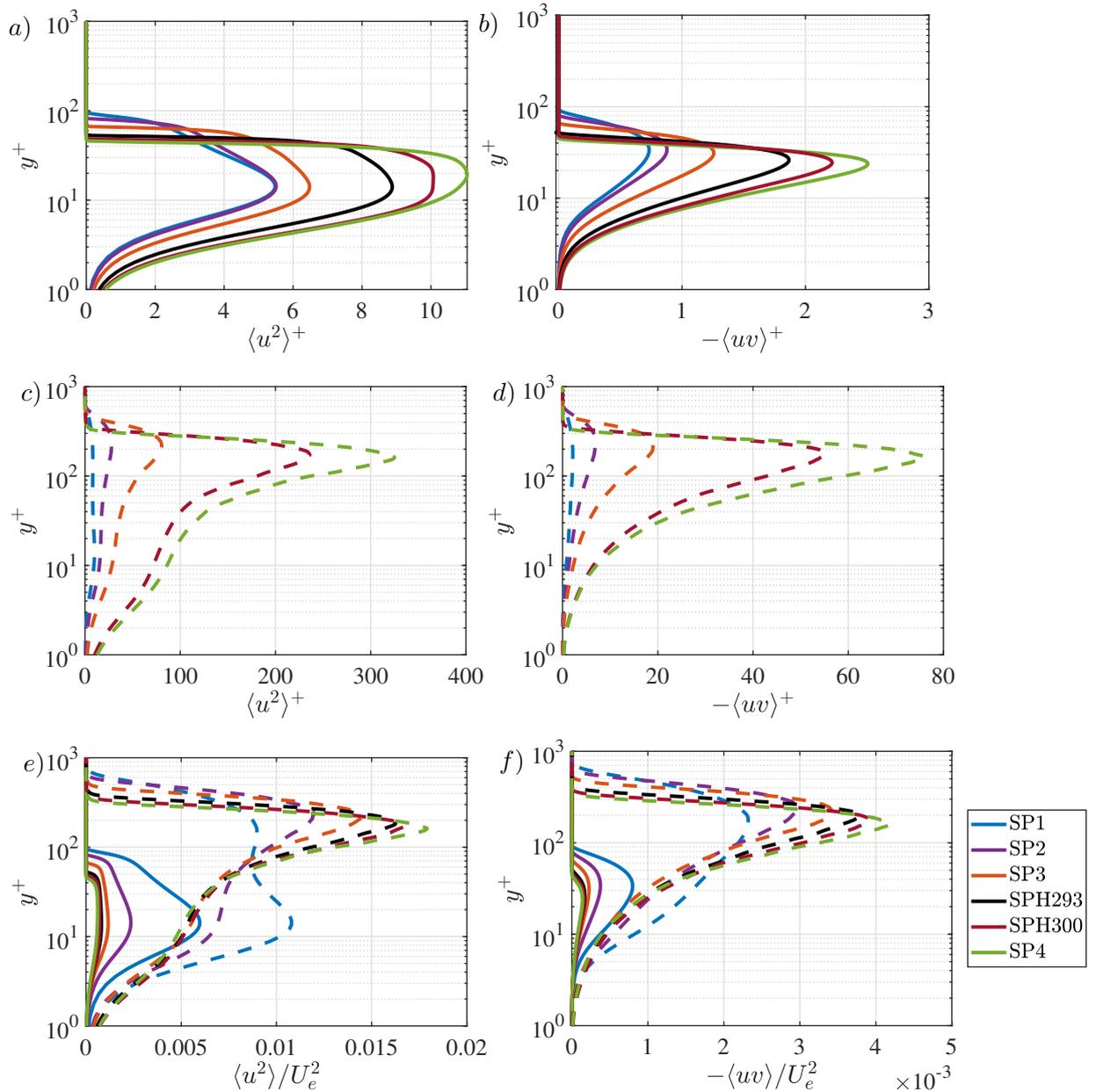

	\section{Results}
	
	\subsection{The wall-normal distribution of Reynolds stresses}
	
	Here we consider the streamwise and tangential component of the Reynolds stresses. Figure \ref{rs_profiles}($a,b,c,d$) presents the inner-scaled Reynolds stress profiles as a function of $y^+$ for the original and manipulated cases. In addition, outer-scaled Reynolds-stress profiles are also given in figure \ref{rs_profiles}($e,f$) so that the differences between the original and manipulated cases can be seen more clearly. The profiles show that the elimination of turbulence activity in the outer layer is performed successfully. The boundary where the turbulence is removed is different in all stations because its location is chosen based on local $\delta$ and not the friction-viscous length scale. These results show that the inner-layer turbulence can sustain itself without any effect from  outer turbulence. This does not necessarily mean that the canonical-flow near-wall cycle exists in the near-wall region. But it nonetheless shows that there is a mechanism sustaining the near-wall turbulence.

	One interesting result is that the Reynolds stress levels have the same order of magnitude in the manipulated case when normalized with friction-viscous scales, as seen in figure \ref{rs_profiles}($a,b$). In contrast, in the original case, figure \ref{rs_profiles}($c,d$) shows that friction-viscous scales fail to scale Reynolds stresses even for moderate velocity defect cases. This result suggests that although the pressure force may affect the production of inner turbulence for large-defect TBLs, it is still the mean shear that drives it. But the situation would necessarily change for even larger defect cases, those very close to or at separation, where mean shear is almost null near the wall. It is rather difficult to compare maxima in the original flow because there is no inner peak except for the first position. However, a rough comparison between SP1o and SP2o, as shown in figure \ref{rs_profiles}($c,d$), shows that the levels of $\langle u^2\rangle^+$ increase from approximately $10$ at SP1 to $16$ at the plateau of SP2 whereas the levels are almost the same in SP1m and SP2m. This change of order of magnitude reflects the fact that the wall-normal transport of outer turbulence to the near-wall region is important for large-defect APG TBLs.
	
{ Besides the Reynolds stress levels, the location of the inner peak of the $\langle u^2\rangle^+$ profiles is the same in the first three positions of the manipulated case, $y^+\approx 13$. Furthermore, the $\langle u^2\rangle^+$ profiles have similar shapes up to SP3, but start changing further downstream as the defect increases. To further examine this, we consider two additional streamwise positions: SPH293 and SPH300 where $H$ is $2.93$ and $2.97$, and $\beta_i$ is $0.38$ and $0.50$, respectively. It is clear that the profile keeps its shape at SPH293 but then a plateau replaces the peak at SPH300. A maximum appears again at $y^+\approx 25$ at SP4. A similar trend of change exists for the $\langle uv \rangle^+$ and production of $\langle u^2\rangle$ (not given here) profiles too. The peak changes from $y^+\approx 35$ to $24$ with increasing defect for $\langle uv \rangle$. As for the production, the production peak is located at $y^+\approx10$ for the first three positions of the manipulated case, but it is at $y^+\approx25$ in SP4 as consistent with the Reynolds-shear stresses. It is not clear if these differences in profiles stem from the fact that the turbulent/non-turbulent interface gets closer to the maximum of $\langle u^2\rangle$ as the defect increases, thereby maybe influencing the turbulent structures.

	\begin{figure}[h!]
		\centering
				\includegraphics[scale=0.5]{ 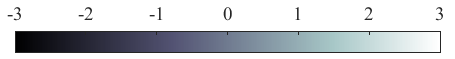}		
				
		\includegraphics[clip,trim=0cm 1.8cm 1cm 1cm ,scale=0.5]{ 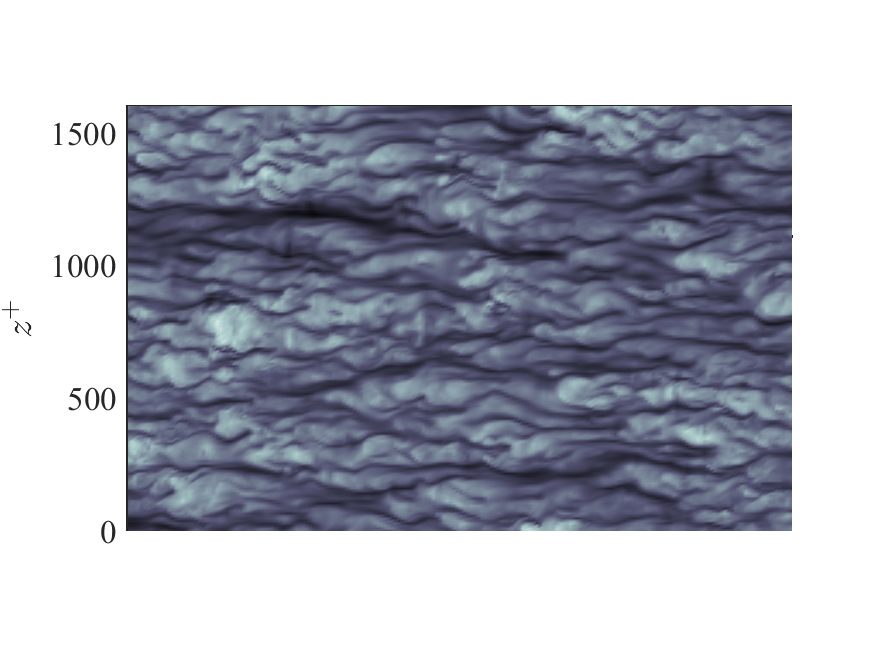}
		\includegraphics[clip,trim=0cm 1.8cm 1cm 1cm ,scale=0.5]{ 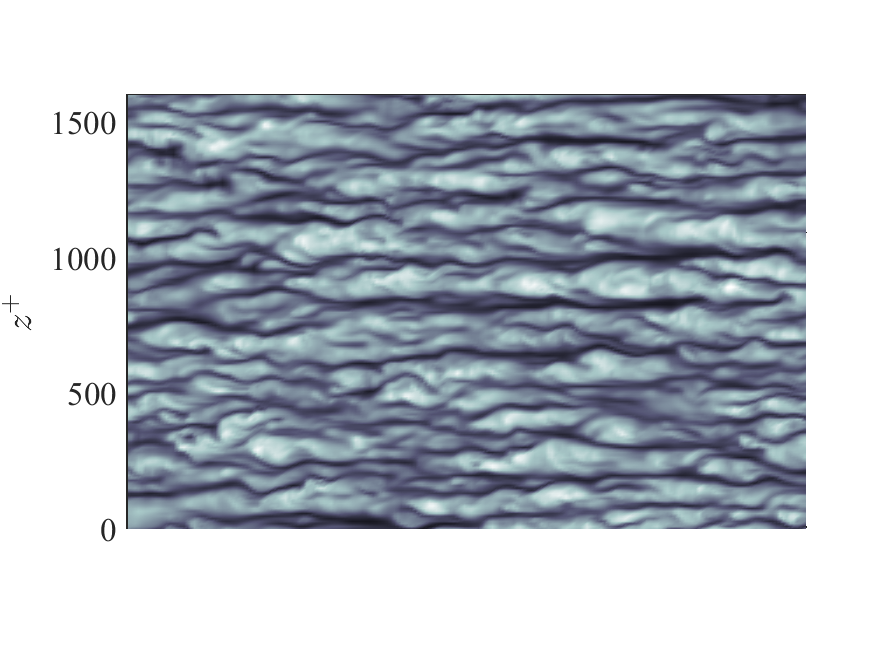}

\includegraphics[clip,trim=0cm 1.8cm 1cm 1cm ,scale=0.5]{ 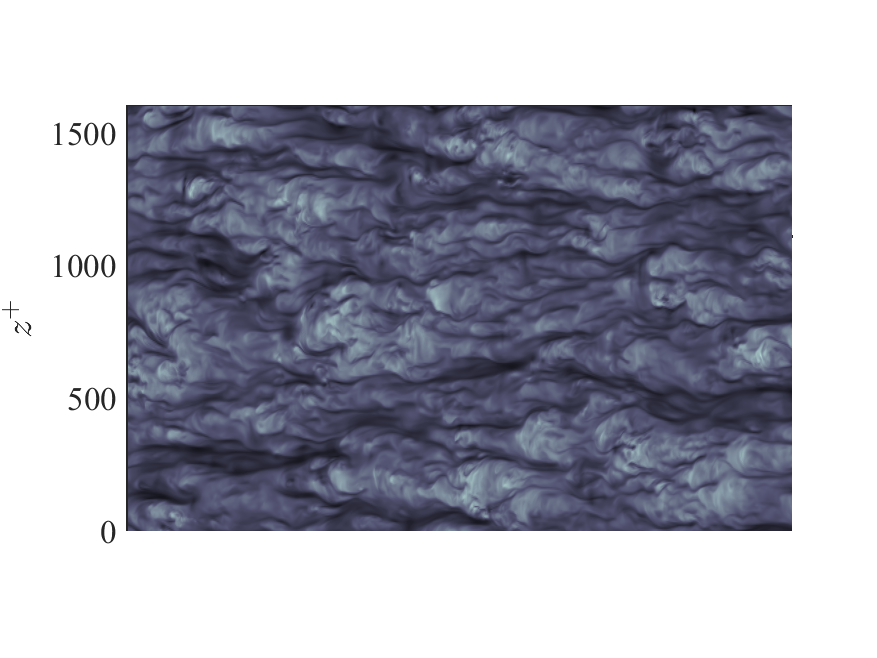} \includegraphics[clip,trim=0cm 1.8cm 1cm 1cm ,scale=0.5]{ 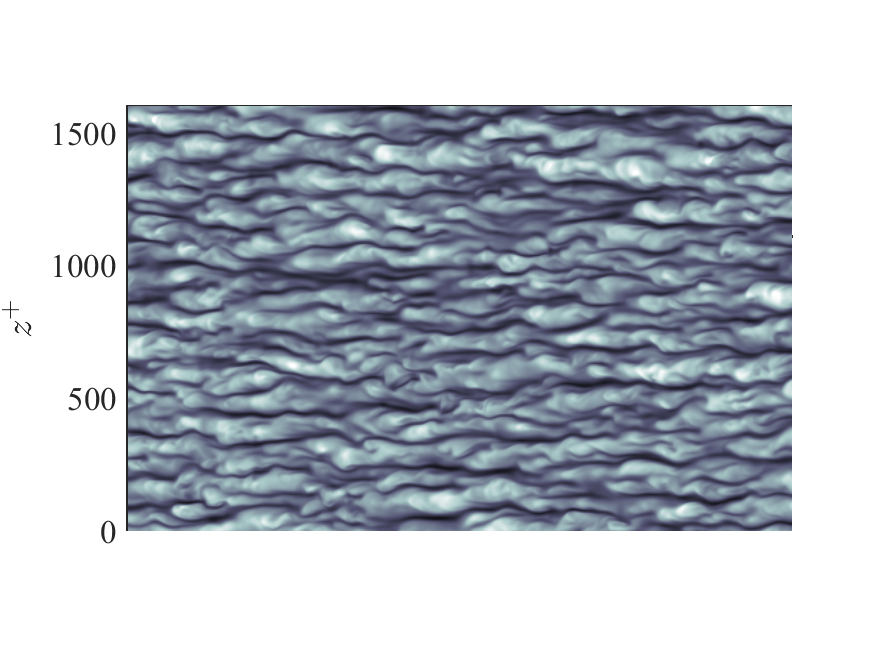}
		
		\includegraphics[clip,trim=0cm 1.8cm 1cm 1cm ,scale=0.5]{ 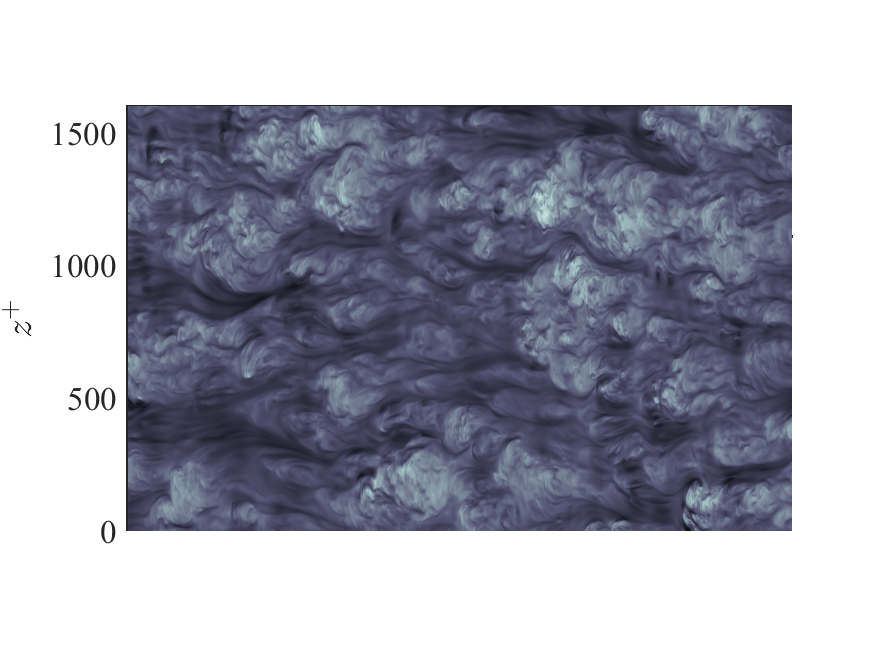} 
		\includegraphics[clip,trim=0cm 1.8cm 1cm 1cm ,scale=0.5]{ 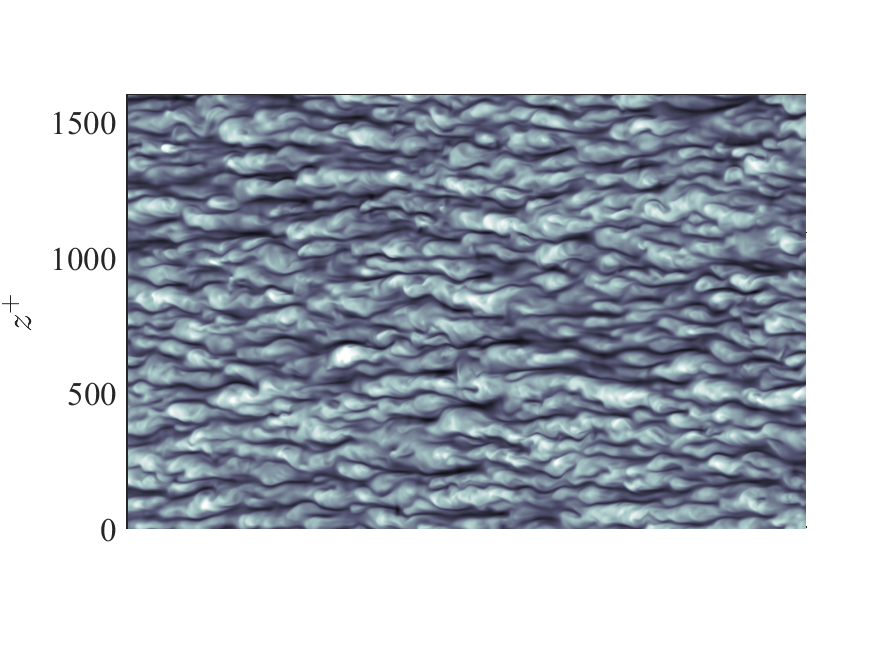}
		
		\includegraphics[clip,trim=0cm 0cm 1cm 1cm ,scale=0.5]{ 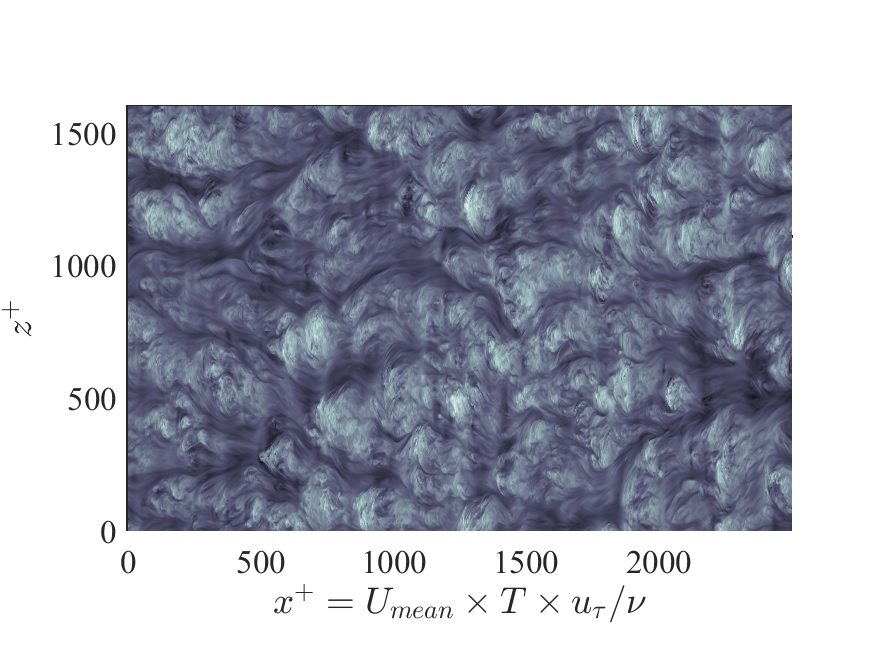} 
		\includegraphics[clip,trim=0cm 0cm 1cm 1cm ,scale=0.5]{ 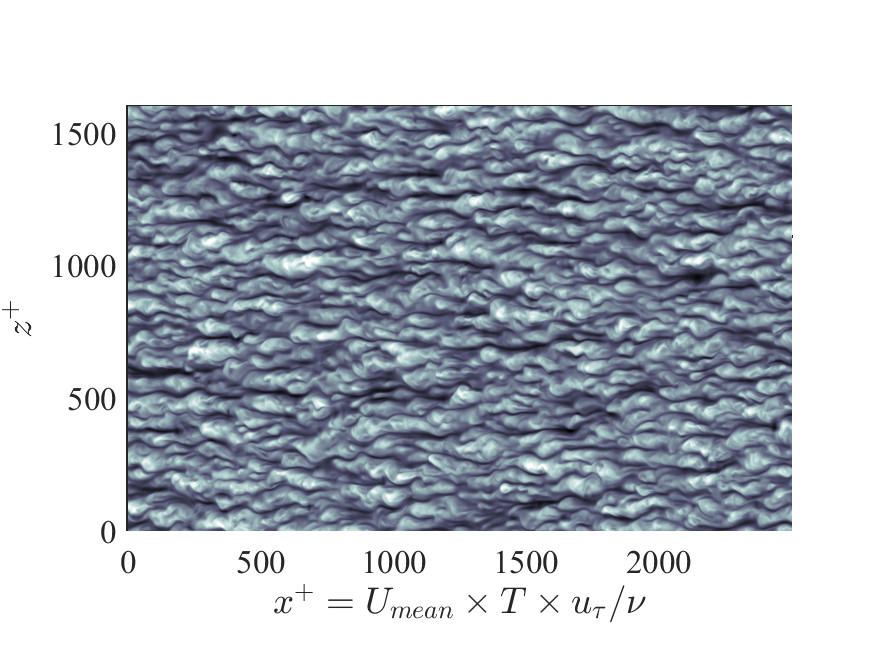}

		\caption{The spatio-temporal evolution of the streamwise velocity fluctuations at $y^+\approx 13$ for the four streamwise positions of both cases as a function of time and $z$. The rows are for, from top to bottom, SP1, SP2, SP3, and SP4. The columns are for the original (left) and manipulated (right) cases. The time axis is transformed into $x$ through Taylor's frozen turbulence hypothesis using the local mean velocity. Axes are normalized with friction-viscous scales. The fluctuations are normalized with the standard deviation of the fluctuations.  }  
		\label{spatio-temp}
	\end{figure}

	\subsection{Spatio-temporal velocity fields}
		Figure \ref{spatio-temp} shows the spatio-temporal evolution of streamwise velocity fluctuations at $y^+\approx13$ for all streamwise positions of the original and manipulated cases. We employ the temporal data that we collected at these positions and transform the data into a spatial representation using Taylor's frozen turbulence hypothesis with the local mean velocity. This way is preferred over the temporal representation because it is convenient to compare the physical length of structures. It is also important to state that the figures do not show the whole spanwise extent for the first three streamwise positions.

	The spatio-temporal evolution in the original case shows that the flow changes considerably from SP1 to SP4. The structures' spanwise width increases significantly with increasing defect. Moreover, they become less streaky and more meandering, especially at SP3 and SP4. In contrast with the original case, the flow field does not change much in the manipulated case. The structures are more compact in large-defect cases, a consequence of using friction-viscous scales, but they are streaky regardless of the defect. As for comparing both cases, the flow fields are similar in the first position. This is expected considering that outer layer turbulence is still weak at SP1o. However, the flow fields become different from each other progressively as the defect increases from SP1 to SP4. In the last two positions, SP3 and SP4, the flow fields are completely different from each other. Neither the structures' features nor their size are similar. The massive difference at SP3/4 is expected because it was noted that the outer layer structures dominate the flow in large-defect APG TBLs \cite{gungor2022energy}. What we are seeing in the original flow are the footprints of outer layer turbulence.

	\subsection{Two-point correlations}
	
	The two-point correlations are examined to investigate the spatial features of $\langle u^2\rangle$- and $\langle uv\rangle$-carrying structures in the inner layer of APG TBLs. We compare the original and manipulated cases for the four streamwise positions, as described earlier. The two-point cross-correlation coefficient for two generic variables is defined as follows,

	\begin{equation}
		C_{ {ab}} (\emph{\textbf{r}},{\emph{\textbf{r}}'}) = \frac{\langle a(\emph{\textbf{r}})b({\emph{\textbf{r}}'}) \rangle}{ \sigma_a(\emph{\textbf{r}})\sigma_b({\emph{\textbf{r}}}') }
	\end{equation}

	\begin{figure}
		\centering
		
		\includegraphics[scale=0.45]{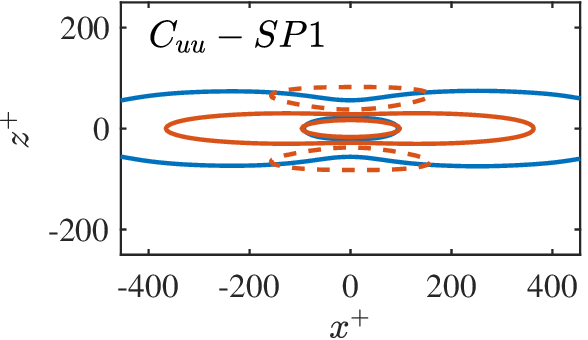}
		\includegraphics[scale=0.45]{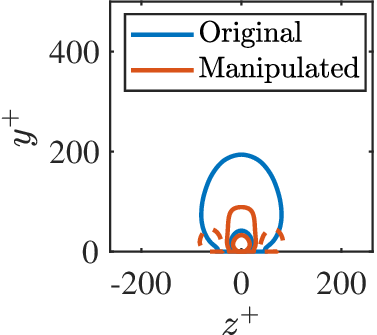}
		\includegraphics[scale=0.45]{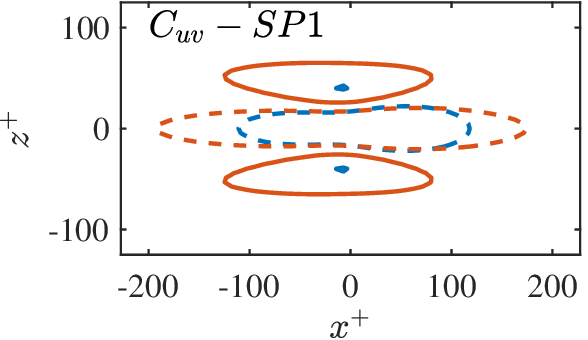}
		\includegraphics[scale=0.45]{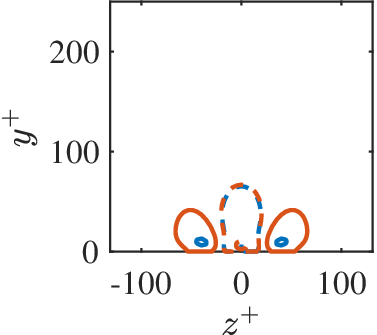}
		
		\includegraphics[scale=0.45]{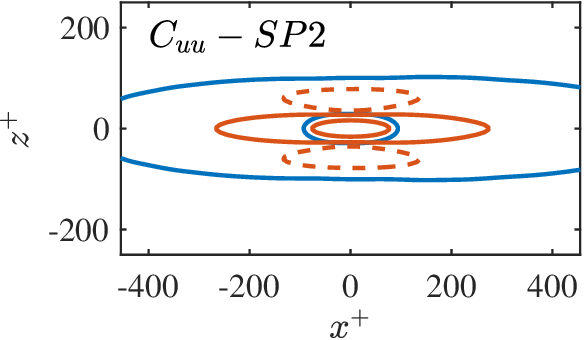}
		\includegraphics[scale=0.45]{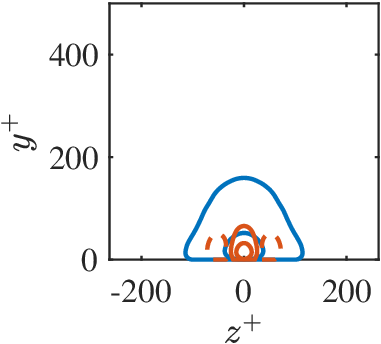}
		\includegraphics[scale=0.45]{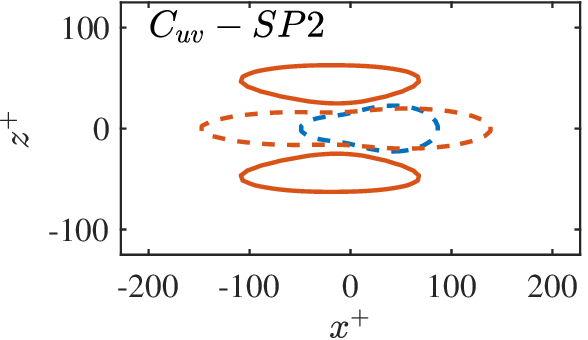}
		\includegraphics[scale=0.45]{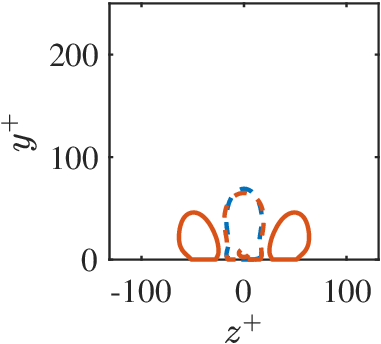}
		
		\includegraphics[scale=0.45]{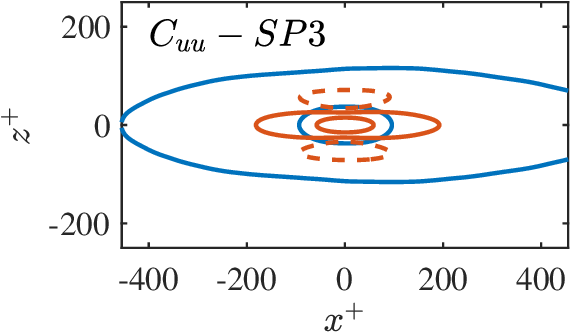}
		\includegraphics[scale=0.45]{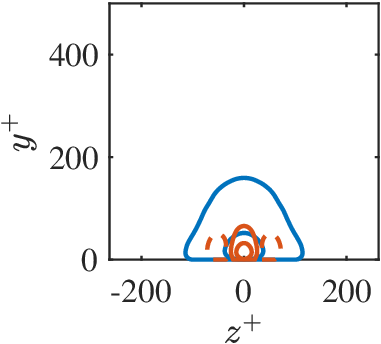}
		\includegraphics[scale=0.45]{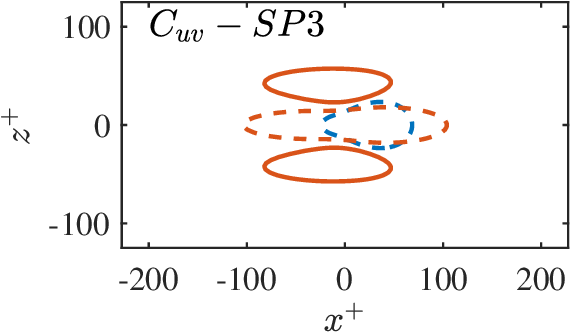}
		\includegraphics[scale=0.45]{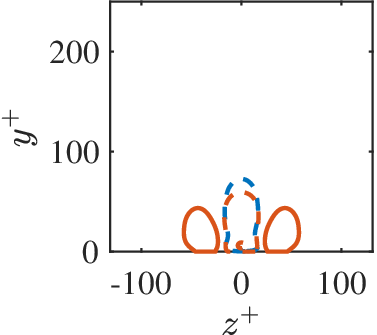}
		
		\includegraphics[scale=0.45]{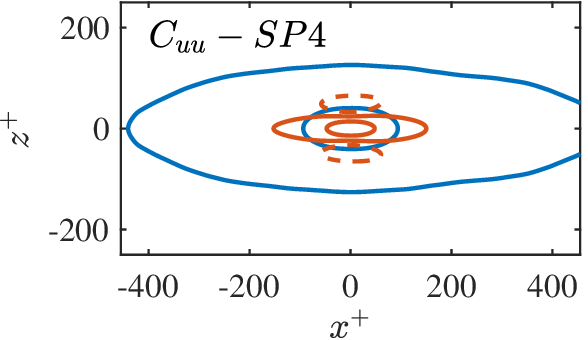}
		\includegraphics[scale=0.45]{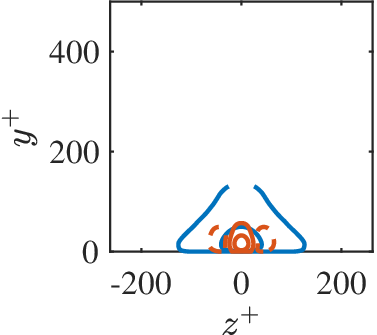}
		\includegraphics[scale=0.45]{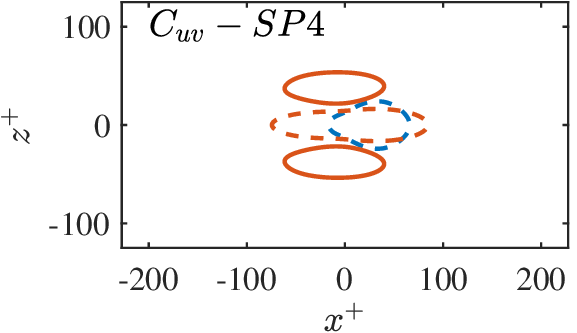}
		\includegraphics[scale=0.45]{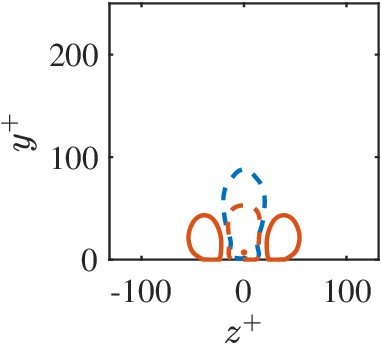}	
			
		\caption{The two-point correlations $C_{uu}$ (two columns at the left) and $C_{uv}$  (two columns at the right) at $y^+\approx13$. The contour levels are for 0.5 and 0.1 for straight lines, and -0.1 for the dashed lines. The rows are for SP1, SP2, SP3, and SP4.  }
		\label{corr_fig}
	\end{figure}

	\noindent where $\emph{\textbf{r}}$ is the reference point, $\emph{\textbf{r}}'$ is the moving point, $a$ and $b$ are generic variables, and $\sigma$ is the standard deviation. For $\langle u^2\rangle$- and $\langle uv\rangle$-carrying structures, we consider the auto-correlation of the streamwise component ($C_{uu}$) and the cross-correlation of $u$ and $v$ ($C_{uv}$). Figure \ref{corr_fig} presents $C_{uu}$ and $C_{uv}$ for two cases at $y^+\approx13$, where the contour levels are 0.5, 0.1, and -0.1. It should be noted that the two-point correlation does not necessarily indicate the size or shape of the structures \cite{hutchins2007evidence}, but low-level contours are associated with large-scale structures.

	The correlation contours are considerably different between the original and manipulated cases for $C_{uu}$ and the difference is much more pronounced in large-defect cases, consistent with the spatio-temporal distributions of figure 4. The contours are both streamwise elongated at SP1 as seen in the $xz$-plane. The high-level contour (0.5) is almost identical in both cases. However, the low-level contours differ considerably from one case to the other. The extent of the contours is much larger in the original case than in the manipulated case. Both high and low-level contours become wider in the original case with increasing velocity defect. Furthermore, they are no longer streamwise elongated and the low-level contours lose their distinctive wasp shape that is narrow at the center. The correlation contours' shape becomes an oval. In contrast, the contours remain almost identical, with more minor variations in their size in the manipulated case. 
	
	The dramatic differences between the two cases exist in the $zy$-plane too. The shape of the contours is completely different in both cases, even at SP1 which is not the case for the contours on the $xz$-plane. As the defect increases, the contours of the original case change significantly and adopt a triangle-like shape where the contours are broader below the reference point and narrower above. This was already reported before for the same database for SP3 \cite{gungor2023coherent}. But here we also consider SP4 which has a considerably higher velocity defect.  This triangle-like shape does not exist in small-defect APG TBLs or ZPG TBLs as reported before. Despite these major changes in the original case, the contours remain almost identical and conserve their distinctive shape in the manipulated case. They are completely different in both cases regardless of the streamwise positions. Furthermore, as is the $xz$-plane, the contours are much more compact in the manipulated case than in the original case.

	In addition to the shape and size of the contours, the other major difference, maybe the most important one, is the presence of negative contours at both sides of the positive contours in the spanwise direction in the manipulated case. This spatial organization of the contours indicates the low- and high-speed streak couples. The two-point correlations show that the streak couples exist from SP1 to SP4. The streak couples have the expected distance of $\Delta z^+\approx 50$ at SP1 but this distance diminishes at large defect cases.  That streak couple exist regardless of the defect is an important finding since it shows that the streaks and therefore the near-wall cycle may exist even in flows with very-large velocity defect, where the mean shear is considerably lower than the one in canonical flows. 
	
	It should be noted that the lack of positive-negative contour couples in the original flow case does not indicate that the streak pairs are absent, because outer turbulence leaves its signature in the two-point correlations in this case. The positive-negative contour couples do not exist for ZPG TBL cases either \cite{gungor2023coherent}. However, it is well-known that the streaks and the near-wall cycle exist in the inner layer of canonical flows.

	The reason for the differences in the size and shape of the contours between the two cases, as described above, is probably the interference of outer-layer structures in the inner layer. As discussed earlier, the outer layer turbulence becomes dominant with increasing velocity defect as figure \ref{rs_profiles} shows. Due to this, the imprints of large-scale outer-layer structures probably become more relevant in the inner layer in large-defect cases and therefore modify more importantly the coherency in the inner layer.

	The differences for the cross-correlation of $u$ and $v$ between the two cases are milder than for $C_{uu}$. We start by discussing the meaning of cross-correlations because it is not as straightforward as auto-correlations. Sillero et al. \cite{sillero2014two} stated that $C_{uv}$ is more like a shorter version of $C_{uu}$ and $C_{vu}$ is like a longer version of $C_{vv}$. This means the cross-correlation does not represent the $\langle uv \rangle$-carrying structures. The best interpretation of $C_{uv}$, as Sillero et al. \cite{sillero2014two} pointed out, is the extent of a streak associated with a sweep or ejection. This is even more useful for our analysis because it directly features the streaks, ejections and sweeps that are playing a role in the turbulence production while $C_{uu}$ indicates the energy-carrying structures. The cross-correlation contours have the negative-positive couples for all defect cases in the manipulated case like the auto-correlation contours and retain some of its features.

As discussed above, the correlation contours conserve their shape in the manipulated case to some extent. Because of this, we can check the scaling of the correlation contours. Figure \ref{scaling} presents $C_{uu}$ and $C_{uv}$ of the manipulated case with normalized using the friction-viscous length scale ($l^+=\nu/u_\tau$) and pressure-viscous length scale ($l^{pi} = \nu/u^{pi}$, see equation \ref{upi} for $u^{pi}$ where $p$ is the pressure.). 

\begin{equation}
 u^{pi} =  \bigg ( \frac{\nu}{\rho}\frac{dp}{dx} \bigg )^{1/3}
	\label{upi}\end{equation}

\begin{figure}
	\centering
	
	\hspace{-.2cm}	\includegraphics[scale=0.35]{ 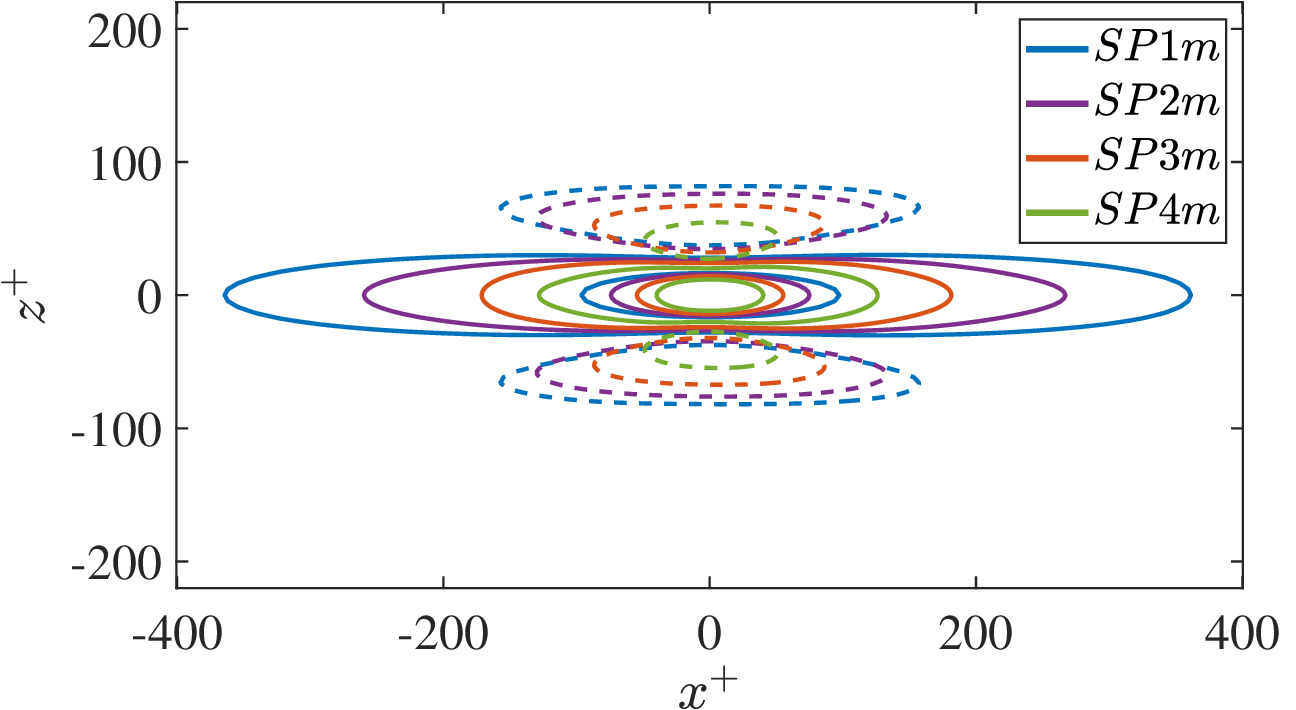}
	\hspace{-0.2cm}		\includegraphics[scale=0.35]{ 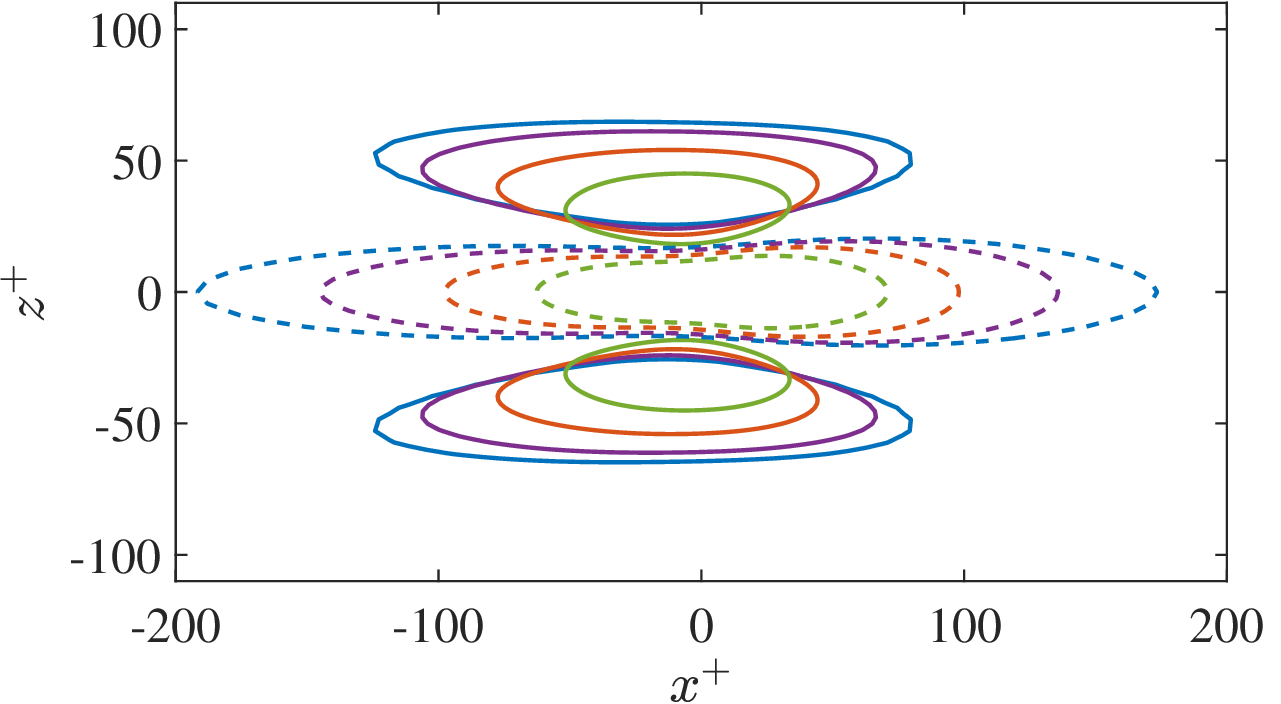}

	\includegraphics[scale=0.35]{ 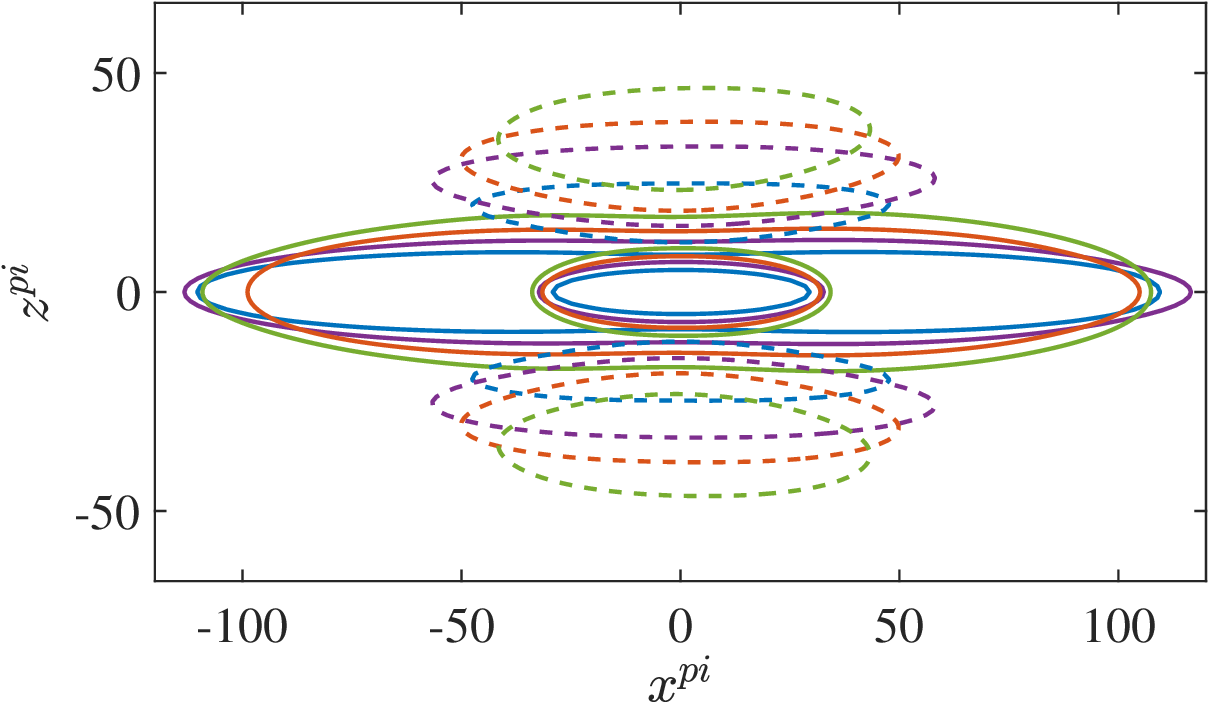}
	\hspace{.3cm}			\includegraphics[scale=0.35]{ 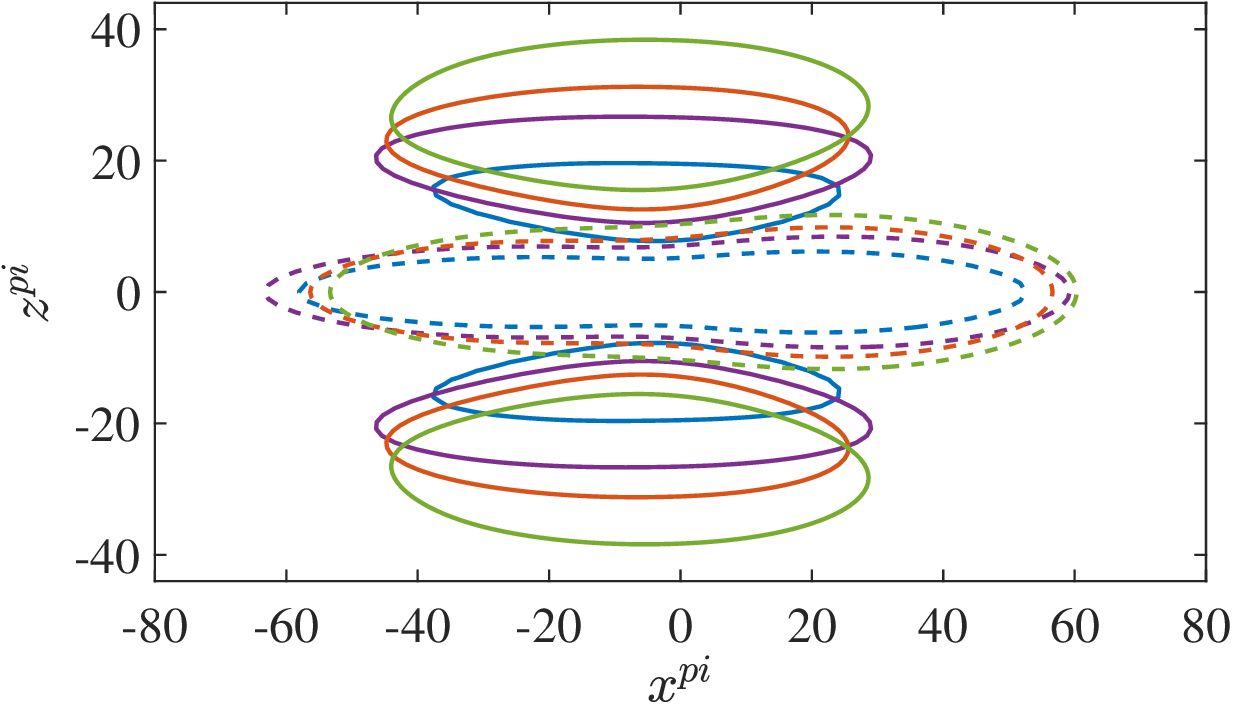}
	\caption{The two-point correlations $C_{uu}$ (left) and $C_{uv}$ (right) at $y^+\approx13$. Axes are normalized with friction-viscous (top) and pressure-viscous (bottom) units. The contour levels are for 0.5 and 0.1 for straight lines, and -0.1 for the dashed lines.}
	\label{scaling}
\end{figure}

The correlation contours of both $C_{uu}$ and $C_{uv}$ become smaller with increasing velocity defect when the axes are normalized with friction-viscous units. This is expected since $u_\tau$ is no longer a valid velocity scale for large-velocity defect cases (it is zero at separation). This is true even if we saw earlier on that friction-viscous scales can scale the position of the maximum of $\langle u^2\rangle$ at least up to the large-defect case of SP4, because these are two different scaling considerations. On the other hand, pressure-viscous units scale the streamwise extent of the contours for $C_{uu}$ and $C_{uv}$ very well. It shows that the structure's streamwise length is affected by the pressure gradient. But such a scaling cannot be used for ZPG TBLs since $u^{pi}=0$.

	\subsection{Spectral distributions of energy, production and inter-component energy transfer }
	
	To further investigate the presence of the near-wall cycle, we examine energy transfer mechanisms through the spectral distributions of energy, production and pressure strain. Production is the extraction of energy from the mean flow to $\langle u^2\rangle$ and the pressure strain governs the inter-component energy transfer between the three normal components. Here, to simplify the discussion, we only consider the spectra related to $\langle u^2\rangle$. For analyzing the energy transfer mechanisms, we employ the transport equations for the two-point correlations tensor as explained in Gungor et al. \cite{gungor2022energy}. Figure \ref{spectra} presents the spectral distributions of the three streamwise positions (SP2 is not given) at $y^+\approx13$ along with the ones for the channel flow of Lee \& Moser \cite{lee2019spectral}, which is included for comparing both APG TBL cases with canonical flows.

The spectral distributions of energy, production and pressure-strain in the channel flow and SP1o are very similar to each other as reported before \cite{gungor2022energy}. The minor difference is that the energy spectra at SP1o have a tail which is the signature of the imprints of the outer-layer structures. Regarding the production and pressure strain, they are also very similar, despite the minor differences in their size. Production structures are slightly shorter and narrower than the most energetic structures in both cases. As the defect increases from SP1 to SP4, the energy spectra become completely different and the energy resides at very large structures, the size of which is similar to those in the outer layer. The difference in production and pressure strain is not as dramatic as the difference in the energy spectra. The structures' aspect ratio remains relatively close. Furthermore, the pressure strain structures are still shorter than the production structures while having a very similar spanwise width. Despite these minor similarities, the shape of the spectra of production and pressure strain at SP3 and SP4 has changed.

In the small-defect case of the manipulated case, the spectra of production, pressure strain and energy are very similar to the ones in channel flow and SP1o. Interestingly, the channel flow case and SP1m are more similar to each other than SP1o and SP1m. For instance, the energy spectra at SP1m do not have the tail at large wavelengths of SP1o. This shows the effect of elevated outer-layer activity on the inner-layer even if SP1 is a small-defect APG case.

\begin{figure}
	\centering
	
	\includegraphics[scale=0.52]{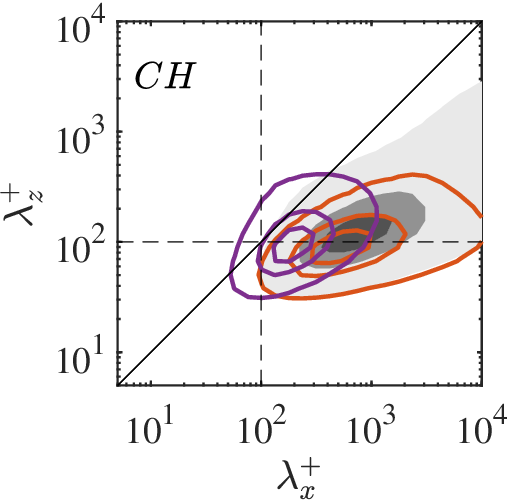}
	\includegraphics[scale=0.52]{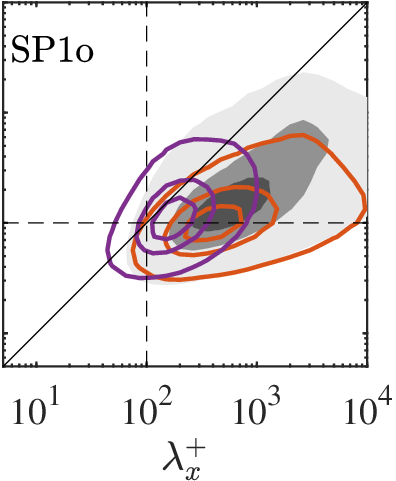}
	\includegraphics[scale=0.52]{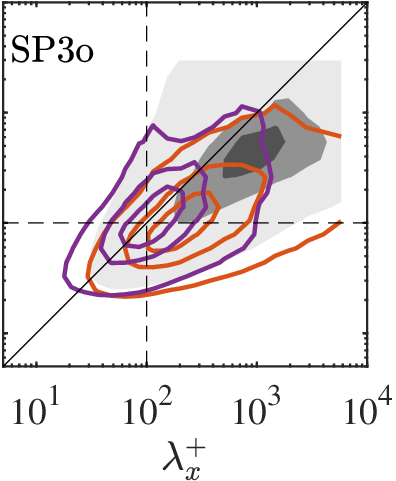}
	\includegraphics[scale=0.52]{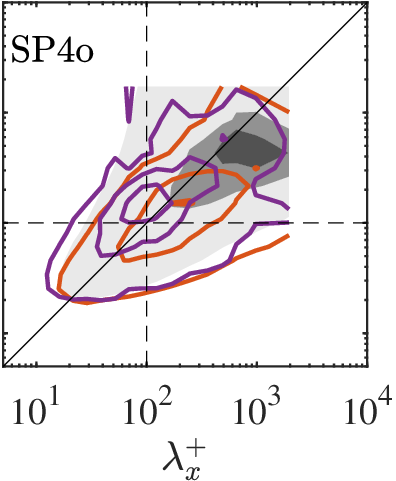}

	\hspace*{3.6cm}	\includegraphics[scale=0.52]{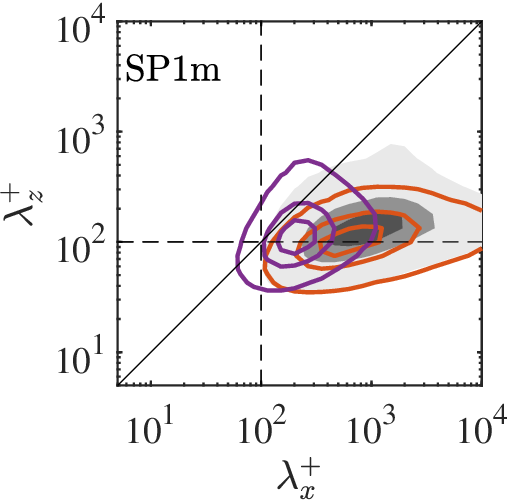}
	\includegraphics[scale=0.52]{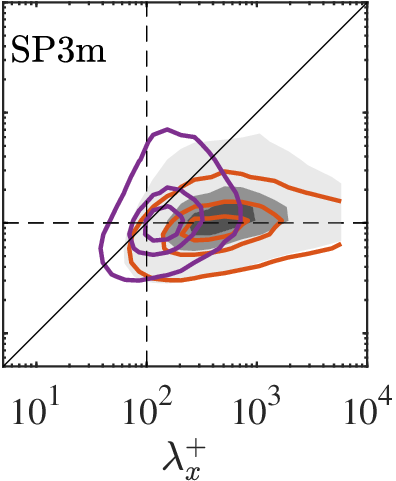}
	\includegraphics[scale=0.52]{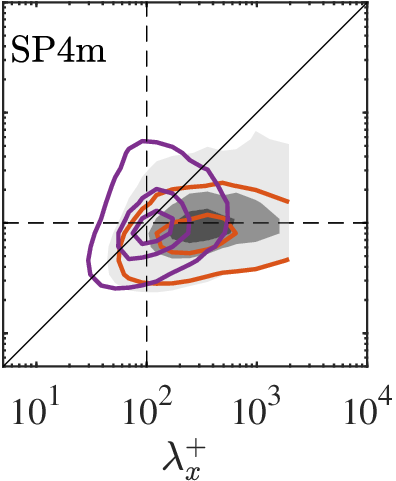}
	
	\caption{The spectral distibutions of energy (shaded), production (red), and pressure strain (purple) at $y^+\approx 13$. Top row from left to right: The channel flow case of Lee and Moser \cite{lee2019spectral}, SP1o, SP3o, and SP4o. The bottom row from left to right: SP1m, SP3m, and SP4m. The levels are [0.1 0.5 0.8] of maxima of spectra.}
	\label{spectra}
\end{figure}

In contrast with the original case, the spectral distributions remain very similar in the manipulated case as the defect increases from SP1 to SP4. The shape of the distributions, the relative size and the aspect ratio of structures are all almost identical.  There is a minor shift towards the smaller wavelengths with increasing defect but this is, at least in part, because friction-viscous scales are used. The difference between the original and manipulated cases at large-defect is massive. It is much more pronounced than the small-defect case. Even though the inner contours of production and pressure-strain are at similar wavelengths (for instance, $\lambda_z^+$ of the pressure-strain peak is 100 at SP1o and 150 for SP1m.), there is an intense energy and energy transfer at wavelengths that are associated with very-large structures in SP3/4. This is expected considering the outer flow dominates the inner layer in SP3/4 and those large-scale structures are probably the imprints of large-scale outer-layer structures.

The most important outcome of the similarity of production and pressure-strain structures in the manipulated case as discussed above is that energy transfer mechanisms are probably the same or very similar at all streamwise positions. This means the near-wall cycle probably exists even if the defect is as large as at SP4, where $\beta_i$ and $H$ are approximately $0.62$ and $3.11$. It also indicates that the near-wall cycle may survive even when the mean shear is considerably lower than the one in canonical flows.

One interesting point is that energy spectra have a clear peak and the energy resides at wavelengths that are associated with the inner-layer streaks regardless of the velocity defect. It is still not clear if these streaks are convected from the upstream or generated locally. However, the fact that the spectra of production and pressure-strain are also very similar suggests the latter.

	\section{Conclusion}
	
	This study examines the turbulence in the near-wall region of APG TBLs to further understand the effect of outer-layer turbulence on inner-layer turbulence activity and dynamics in low shear environment. For this, we employ two non-equilibrium APG TBLs. The first one is the non-equilibrium APG TBL of Gungor et al. \cite{gungor2022energy}. The other one is a novel APG TBL database that is identical to the first one with a major change: the outer layer turbulence is artificially removed. The major findings of this study are listed below.

	\begin{enumerate}

		\item The inner-layer turbulence sustains itself regardless of the mean shear in the inner-layer when the outer layer does not exist. 
		
		\item The streak couples continue to exist in the near-wall region even when the flow is very close to separation in the manipulated flow whereas they cease to exist in large-defect cases of the original flow.
		
		\item It is important to state that it is not clear if these streaks in the near-wall region of the manipulated flow are generated locally or simply convected from upstream where the shear is still high. 
		
		\item The pressure-viscous length scales the streamwise extent of $C_{uu}$ and $C_{uv}$ contours while friction-viscous length does not. Despite this, neither friction-viscous nor pressure-viscous length scale the spanwise extent of the contours.

		\item The spectral characteristics of energy, production and pressure-strain are very similar to those of canonical wall flows in the inner layer regardless of the velocity defect when the outer layer is removed.
		
		\item This suggests that energy transfer mechanisms such as production or inter-component energy are similar in the inner layer and therefore the near-wall cycle may still exist even when the flow is very close to separation.

	\end{enumerate}

\noindent Our analysis shows that the streaks and the near-wall cycle exist in the large-defect APG TBLs when the outer layer is removed. However, it is not clear if the structures playing a role in the near-wall cycle co-exist with the imprints of large-scale outer-layer structures or simply vanish. If they do co-exist, then how do they interact with the imprints of large-scale structure? More work is required to further investigate this. 

	\section{Acknowledgments}

This work was supported in part by the European Research Council under the Caust grant ERC-AdG-101018287. The authors would like to thank Prof. Javier Jim\'enez for organizing the Fifth Madrid Turbulence Workshop. The computational resources were provided by Calcul Qu\'ebec (www.calculquebec.ca) and the Digital Research Alliance of Canada (www.alliancecan.ca). This  work  was  also  funded  in  part by NSERC of Canada.

	\section*{References}
	\bibliographystyle{iopart-num}
	\bibliography{references}
	
\end{document}